\documentclass[prd,superscriptaddress,twocolumn,showpacs,amsmath]{revtex4-1}
\usepackage{graphicx}
\usepackage{color}
\usepackage{booktabs}
\usepackage{hyperref}

\begin{document}
  \title{Critical behavior of lattice Schwinger model with topological term at $\theta=\pi$ using Grassmann tensor
  renormalization group}

  \author{Yuya Shimizu}
  \affiliation{RIKEN Advanced Institute for Computational Science, Kobe, Hyogo 650-0047, Japan}

  \author{Yoshinobu Kuramashi}
  \affiliation{Faculty of Pure and Applied Sciences, University of Tsukuba, Tsukuba, Ibaraki
    305-8571, Japan}
  \affiliation{Center for Computational Sciences, University of Tsukuba, Tsukuba, Ibaraki
    305-8577, Japan}
  \affiliation{RIKEN Advanced Institute for Computational Science, Kobe, Hyogo 650-0047, Japan}

  \begin{abstract}
    Lattice regularized Schwinger model with a so-called $\theta$ term is studied by using the Grassmann tensor
    renormalization group. We perform the Lee-Yang and Fisher zero analyses in order to investigate the phase
    structure at $\theta=\pi$.
    We find a first order phase transition at larger fermion mass.
    Both of the Lee-Yang zero and Fisher zero analyses indicate that
    the critical endpoint at which the first order phase transition terminates belongs to the Ising
    universality class.
  \end{abstract}
  \pacs{05.10.Cc, 11.15.Ha}
  \date{\today}
  \maketitle

  \section{Introduction}
  \label{sec:intro}
  The Monte Carlo simulation of lattice gauge theory is quite powerful to study nonperturbative
  phenomena of particle physics. However, when the action has an imaginary part like the $\theta$ term,
  it suffers from the numerical sign problem, failure of usual importance sampling techniques.
  The effect of the $\theta$ term on non-Abelian gauge theory, especially quantum chromodynamics (QCD)
  is important, because it is related to a famous unsolved problem, ``strong CP problem''.
  See Ref.~\cite{Vicari:2008jw} for a recent review on gauge theory with the $\theta$ term.
  In order to tackle such a problem, another approach is desired.
  Lattice gauge theory with the $\theta$ term shares the difficulty with finite density lattice QCD.
  Therefore, developing techniques to solve or by-pass the sign problem also leads to a lot of progress
  in the study of the QCD phase diagram at finite temperature and density.

  It is well-known that the $\theta$ term has a non-trivial contribution to Abelian gauge theory in two
  dimensions, also. Coleman argued that the (massive) Schwinger model, 2D QED, undergoes a phase transition
  at $\theta=\pi$ as ${m}/{g}$ increases where $m$ is the fermion mass and $g$ is the coupling
  constant~\cite{Coleman:1976uz}.
  It was followed by numerical lattice calculations and they succeeded in estimating the critical
  endpoint~\cite{Hamer:1982mx,Schiller:1983sj,Byrnes:2002nv}.
  However, all these are based on the Hamiltonian lattice gauge theory and numerical studies with the
  Euclidean lattice gauge theory are falling behind: Up to now only pure lattice gauge
  theory has been studied in the Euclidean formulation because it is
  analytically solvable~\cite{Wiese:1988qz,Hassan:1994wy,Hassan:1995dn,Plefka:1996tz}. 
 Once including fermions, we have not yet established any reliable method which is effective at
  $\theta=\pi$ in the Euclidean formulation. 

  Recently the authors have successfully applied the Grassmann tensor renormalization group (GTRG)~\cite{Gu:2010aa} to the analysis
  on the lattice Schwinger model in the Euclidean formulation~\cite{Shimizu:2014uva}. The GTRG method directly treats the Grassmann numbers without relying on the pseudofermion technique employed in the hybrid Monte Carlo algorithm so that the computational cost is comparable to the bosonic case. Another virtue is that it does not suffer from the sign problem caused by the fermion determinant. 
  In this paper, we extend the GTRG method to the case including the $\theta$ term, where the action becomes complex, and demonstrate that it enables us to
  investigate the phase structure at $\theta=\pi$.

  This paper is organized as follows. We briefly discuss the Schwinger model with the $\theta$ term in the continuum theory and its lattice formulation in
  Sec.~\ref{sec:sch}.
  In Sec.~\ref{sec:num}, our numerical results obtained by the Lee-Yang and Fisher zero analyses are presented.
  Sec.~\ref{sec:sum} is devoted to summary and outlook.

  \section{Schwinger model with $\theta$ term}
  \label{sec:sch}
  \subsection{Continuum theory}
  Let us briefly describe the Schwinger model with the $\theta$ term.
  The Euclidean action is given by
  \begin{equation}
    S=\int\!d^2x\left\{ \Bar{\psi}(\gamma_\mu\partial_\mu +i\gamma_\mu A_\mu+m)\psi
    +\frac{1}{4g^2}F_{\mu\nu}F_{\mu\nu}\right\},
    \label{}
  \end{equation}
  where $\psi$ is a two-component spinor field and $A_\mu$ is a U(1) gauge field. The field strength is 
defined by
  \begin{equation}
    F_{\mu\nu}=\partial_\mu A_\nu-\partial_\nu A_\mu.
    \label{}
  \end{equation}
  Vacua of U(1) gauge theory in two dimensions are labeled by an integer number $Q$ which is computed from
  \begin{equation}
    Q=\frac{1}{4\pi i}\int\!d^2x\,\epsilon_{\mu\nu}F_{\mu\nu},
    \label{}
  \end{equation}
  where $\epsilon_{\mu\nu}$ is an antisymmetric tensor with $\epsilon_{12}=i$. The $\theta$ vacuum is introduced as a
  superposition of the labeled vacua. Therefore, the partition function in the $\theta$ vacuum is expressed
  as
  \begin{align}
    Z&=\sum_{Q=-\infty}^\infty
    e^{i\theta Q}\int\!\mathcal{D}A^{(Q)}\,\mathcal{D}\Bar{\psi}\,\mathcal{D}\psi\,e^{-S[A^{(Q)}]}\\
    &=\sum_{Q=-\infty}^\infty
    \int\!\mathcal{D}A^{(Q)}\,\mathcal{D}\Bar{\psi}\,\mathcal{D}\psi\,e^{-S[A^{(Q)}]+\frac{\theta}{4\pi}
    \int\!d^2x\,\epsilon_{\mu\nu}F_{\mu\nu}^{(Q)}}\\
    &=\int\!\mathcal{D}A\,\mathcal{D}\Bar{\psi}\,\mathcal{D}\psi\,e^{-S[A]+\frac{\theta}{4\pi}
    \int\!d^2x\,\epsilon_{\mu\nu}F_{\mu\nu}},
  \end{align}
  where $\theta$ is the vacuum angle.
  In addition, with the use of a chiral transformation,
  \begin{gather}
    \psi\to e^{-i\frac{\theta}{2}\gamma_5}\,\psi,\\
    \Bar{\psi}\to \Bar{\psi}\,e^{-i\frac{\theta}{2}\gamma_5},
  \end{gather}
  the $\theta$ term is canceled by the anomaly and the mass term is modified:
  \begin{equation}
    \begin{split}
      Z=&\int\!\mathcal{D}A\,\mathcal{D}\Bar{\psi}\,\mathcal{D}\psi\\
      &e^{-\int\!d^2x\left\{\Bar{\psi}(\gamma_\mu\partial_\mu +i\gamma_\mu A_\mu+m\cos\theta
      +im\gamma_5\sin\theta)\psi+\frac{1}{4g^2}F_{\mu\nu}F_{\mu\nu}\right\}}.
    \end{split}
  \end{equation}

\begin{figure}
    \centering
    \includegraphics[width=50mm]{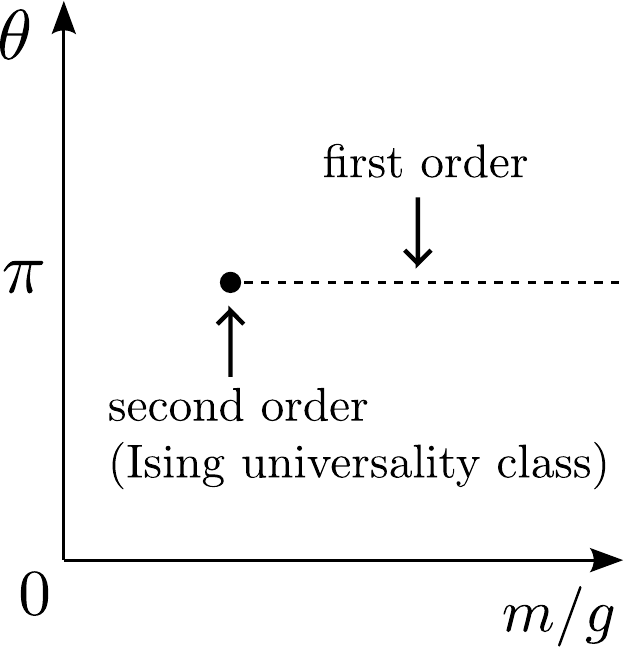}
    \caption{Expected phase diagram of Schwinger model with the 
$\theta$ term. The dotted line denotes a first
    order phase transition, which terminates at a second order phase transition point belonging to the Ising
    universality class.}
    \label{fig:pd}
  \end{figure}

The Schwinger model can be mapped to a bosonic model by using following
  correspondences~\cite{Coleman:1975pw,Coleman:1976uz}:
  \begin{gather}
    S_{m=0}\leftrightarrow \int\!d^2x\left\{\frac{1}{2}(\partial \phi)^2+\frac{g^2}{2\pi}\phi^2\right\},\\
    \Bar{\psi}\psi\leftrightarrow -Cg\cos(2\sqrt{\pi}\phi),\\
    i\Bar{\psi}\gamma_5\psi\leftrightarrow  -Cg\sin(2\sqrt{\pi}\phi),
  \end{gather}
  where $\phi$ is a scalar field and $C$ is some constant which depends on the scheme employed for
  normal-ordering operators~\cite{Coleman:1974bu}. The bosonized version of the partition function is
  \begin{equation}
      Z=\int\!\mathcal{D}\phi\,e^{-\int\!d^2x\left\{\frac{1}{2}(\partial \phi)^2+\frac{g^2}{2\pi}\phi^2
      -Cmg\cos(2\sqrt{\pi}\phi-\theta) \right\}}.
  \end{equation}
  Let's consider the potential term,
  \begin{equation}
    V[\phi]=\frac{g^2}{2\pi}\phi^2-Cmg\cos(2\sqrt{\pi}\phi-\theta).
  \end{equation}
  Intriguing finding is that $\theta=\pi$ is a special case. 
For sufficiently large ${m}/{g}$, $V[\phi]$ becomes a double well potential.
  It tells us that there exists a first order phase transition at the semiclassical level.
  On the other hand, in the limit of ${m}/{g}\to 0$, the second term can be negligible so that $V[\phi]$ has
  an unique minimum. This means that the first order phase transition terminates at some value of $m/g$, where a second
  order phase transition takes place due to the breaking of the Z(2) symmetry. In
  Fig.~\ref{fig:pd} we illustrate the expected phase diagram  of the Schwinger model with the $\theta$ term.
It should be noted that the Ising model has a similar phase structure in the plane of an external magnetic field $H$ and the temperature $T$. In the Ising case, a first order phase transition at lower temperature with $H=0$ terminates at some critical temperature $T_{\rm c}$ where a second order phase transition occurs.  

  \subsection{Lattice formulation}
  We follow the formulation given in Ref.~\cite{Shimizu:2014uva} 
except the additional $\theta$ term.
  Hereafter, all the parameters are expressed by dimensionless quantities multiplied by the lattice spacing
  $a$.

  We employ the Wilson fermion action and plaquette gauge action.
  The Wilson-Dirac matrix $D[U]$ is given by
  \begin{equation}
    \begin{split}
      \Bar{\psi}D[U]\psi=&\frac{1}{2\kappa}\sum_{n,\alpha} \Bar{\psi}_{n,\alpha}\psi_{n,\alpha}\\
      &-\frac{1}{2}\sum_{n,\mu,\alpha,\beta}\Bar{\psi}_{n,\alpha}\{
      (1-\gamma_\mu)_{\alpha,\beta}\,U_{n,\mu}\psi_{n+\Hat{\mu},\beta}\\&+(1+\gamma_\mu)_{\alpha,\beta}
      \,U^\dagger_{n-\Hat{\mu},\mu}\psi_{n-\Hat{\mu},\beta}\},
    \end{split}
  \end{equation}
  where the hopping parameter $\kappa$ satisfies $1/\kappa=2(m+2)$ and an U(1) link
  variable at site $n$ along $\mu$ direction, $U_{n,\mu}$ is related to $A_\mu(n)$ as follows:
  \begin{equation}
    U_{n,\mu}=e^{iaA_\mu(n)}.
    \label{}
  \end{equation}
  $\alpha,\beta$ denote the Dirac indices and $\Hat{\mu}$ represents an unit vector along $\mu$ direction.
  The U(1) gauge action including the $\theta$ term is given by
  \begin{gather}
	  S_g=
	  -\beta\sum_{p}\cos\varphi_p
	  -i\theta Q,\\
          \varphi_p=\varphi_{n,1}+\varphi_{n+\Hat{1},2}-\varphi_{n+\Hat{2},1}-\varphi_{n,2},\\
	  \varphi_{n,1},\varphi_{n+\Hat{1},2},\varphi_{n+\Hat{2},1},\varphi_{n,2}\in [-\pi,\pi],\\
	  Q=\frac{1}{2\pi}\sum_{p}q_p,\\
	  q_p=\varphi_p\quad\text{mod}\ 2\pi
  \end{gather}
  with $\beta=1/g^2$. $\varphi_{n,1},\varphi_{n+\Hat{1},2},\varphi_{n+\Hat{2},1}$ and $\varphi_{n,2}$ are
  phases of U(1) link variables which compose a plaquette variable.
  The range of $q_p$ is $[-\pi,\pi]$ and it can be expressed as follows by introducing an integer $n_p$:
  \begin{equation}
	  q_p=\varphi_p+2\pi n_p,\quad n_p\in\{-2,-1,0,1,2\}.
	  \label{}
  \end{equation}
  For periodic boundary conditions, the topological charge $Q$ should be an integer even on the lattice:
  \begin{equation}
	  Q=\frac{1}{2\pi}\sum_p\varphi_p+\sum_pn_p=\sum_pn_p.
	  \label{}
  \end{equation}
 With the inclusion of the $\theta$ term, the character expansion of the
  Boltzmann weight per plaquette is decomposed as follows~\cite{Hassan:1994wy,Hassan:1995dn}:
  \begin{equation}
    \begin{split}
      &\exp\left\{\beta\cos\varphi_p+i\frac{\theta}{2\pi}q_p\right\}\\
      &=\sum_{m=-\infty}^\infty e^{im\varphi_p}\sum_{l=-\infty}^\infty
      I_l(\beta)\frac{2\sin\left(\frac{\theta+2\pi(m-l)}{2}\right)}{\theta+2\pi(m-l)}\\
      &\simeq\sum_{m=-N_{\rm ce}}^{N_{\rm ce}} e^{im\varphi_p}\sum_{l=-N_{\rm ce}^\prime}^{N_{\rm ce}^\prime}
      I_l(\beta)\frac{2\sin\left(\frac{\theta+2\pi(m-l)}{2}\right)}{\theta+2\pi(m-l)},
    \end{split}
    \label{eq:ce}
  \end{equation}
  where $I_l$ is the modified Bessel function. We choose $N_{\rm ce}$ and $N_{\rm ce}^\prime$ for truncation 
of the summations in the practical numerical calculations. 
This series converges due to rapid decreasing of the modified
  Bessel function with increasing $\vert l\vert$, but
  the rate becomes smaller than the case without the $\theta$ term.

  \section{Numerical analysis}
  \label{sec:num}
  \subsection{Setup}
We perform the Lee-Yang and Fisher zero analyses at $\beta=10.0$ to investigate the phase transition of the model. 
We refer to partition function zeros in the complex $\kappa$ plane as the Fisher zeros in order to distinguish
  them from those in the complex $\theta$ plane which are referred to as the Lee-Yang
  zeros. We employ the GTRG method described in Ref.~\cite{Shimizu:2014uva}, which allows us to estimate partition function zeros. We choose $N_{\rm ce}=20$ and $N_{\rm ce}^\prime=100$ for truncation of the summations in Eq.~\eqref{eq:ce}.
The singular value decomposition in
  the GTRG procedure is truncated with $D=160$. 
We have checked that these choices for $N_{\rm ce}$, $N_{\rm ce}^\prime$ and $D$ provide us sufficiently accurate results for all the parameter sets employed in this work.
  Since the scaling factor of the GTRG transformation is $\sqrt{2}$, we are allowed to evaluate the partition
  function zeros not only at the lattice size
  $L=4,\,8,\,16,\,\cdots$, but also at $L=4\sqrt{2},\,8\sqrt{2},\,16\sqrt{2},\,\cdots$. The periodic boundary
  condition is employed in both directions.

  \subsection{Fisher zero analysis}
  Partition function zeros in a complex parameter plane should approach a phase transition point on the real
  axis as the lattice size $L$ increases. Their scaling behavior, however, depends on what parameter we
  focus on. In case of the hopping parameter $\kappa$, which may correspond 
to the temperature parameter in
  the Ising model, the scaling behavior is governed by the critical exponent for the correlation length $\nu$:
  \begin{gather}
    \text{Re}\,\kappa_0(L)-\text{Re}\,\kappa_0(\infty)\propto L^{-1/\nu},\label{eq:rek}\\
    \text{Im}\,\kappa_0(L)-\text{Im}\,\kappa_0(\infty)\propto L^{-1/\nu},\label{eq:imk}
  \end{gather}
  where $\kappa_0(L)$ denotes the position of a partition function zero in the complex $\kappa$ plane for the
  lattice size $L$. $\text{Re}\,\kappa_0(\infty)$ should agree with the
  critical point $\kappa_c$, while $\text{Im}\,\kappa_0(\infty)$ should be consistent with zero.

  \begin{figure}
    \centering
    \includegraphics[width=80mm]{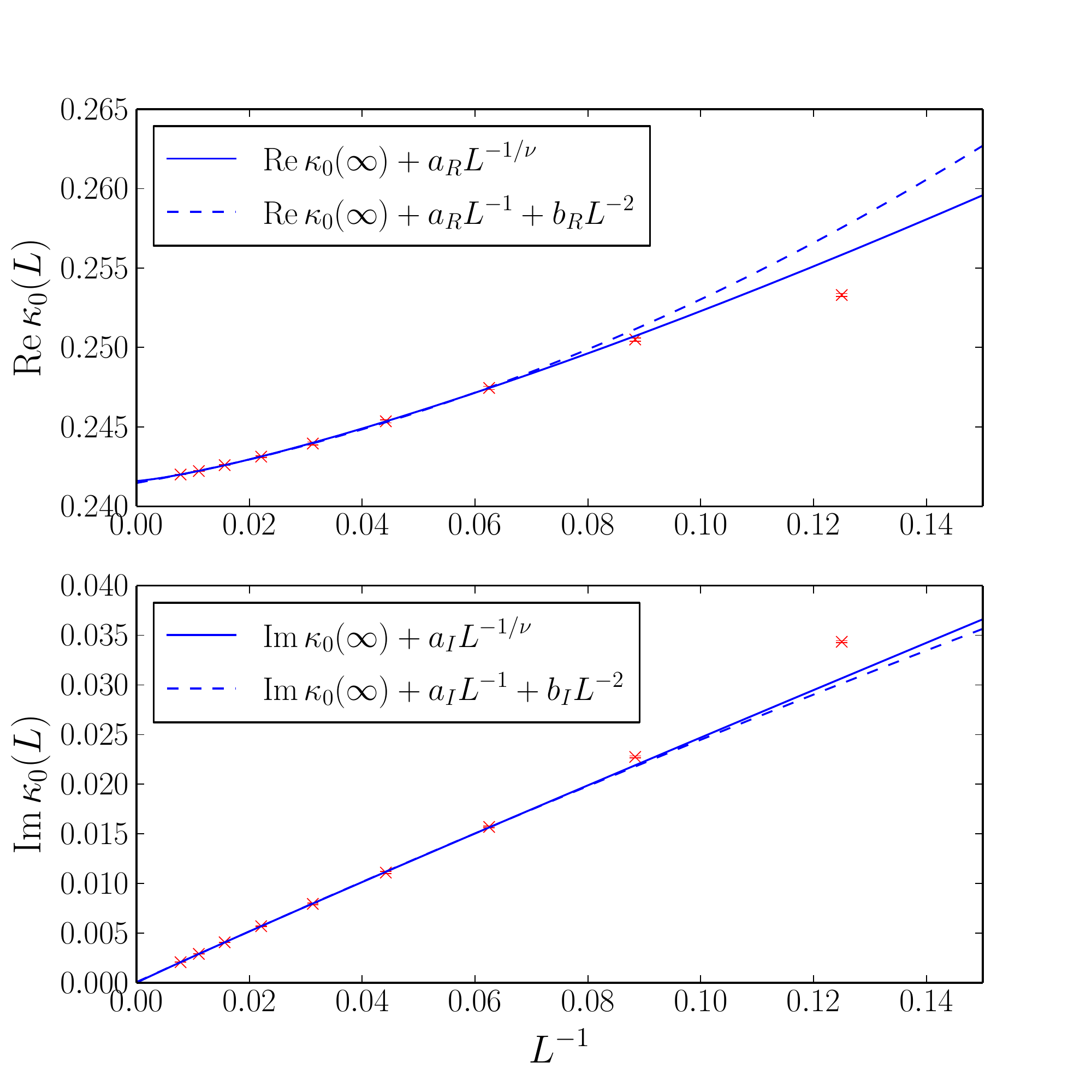}
    \caption{Real (top) and imaginary (bottom) parts of the Fisher zero as a function of $L^{-1}$ at
    $\beta=10.0$. Solid curves represent the fit results with
    $\text{Re}/\text{Im}\,\kappa_0(L)=\text{Re}/\text{Im}\,\kappa_0(\infty)+a_{R/I} L^{-1/\nu}$ and dotted
    ones with Eqs.~\eqref{eq:jfre} and \eqref{eq:jfim}.}
    \label{fig:fisher}
  \end{figure}

  Figure~\ref{fig:fisher} shows finite size scaling plots of both the real and imaginary parts of the
  Fisher zero closest to the real axis.
We locate $\kappa_0(L)$ on the mesh of the discretized $\text{Re}\,\kappa$ and $\text{Im}\,\kappa$ so that
the mesh spacing determines the error bars of $\text{Re}\,\kappa_0(L)$ and $\text{Im}\,\kappa_0(L)$.
  The solid curves denote the fit results with $\text{Re}/\text{Im}\,\kappa_0(L)=\text{Re}/\text{Im}\,
  \kappa_0(\infty)+a_{R/I}L^{-1/\nu}$. The fit range is chosen as $L\in [16,128]$ avoiding possible finite size
effects expected in the small $L$ region. Numerical values for the fit results are listed in Table~\ref{tab:fisher}.
  We observe that the result for $\nu$ in the imaginary part
  is very close to $\nu=1$ which indicates the Ising universality class. On the other hand,
 the real part clearly deviates from $\nu=1$.
  The situation is quite similar to the case without the $\theta$
  term~\cite{Shimizu:2014uva}, where the disagreement can be explained by possible finite size contaminations. 
Let us try the following fit functions with the leading term with $\nu=1$ and the $L^{-2}$ subleading term:
  \begin{gather}
\text{Re}\,\kappa_0(L)-\text{Re}\,\kappa_0(\infty)=a_RL^{-1}+b_RL^{-2},\label{eq:jfre}\\
\text{Im}\,\kappa_0(L)-\text{Im}\,\kappa_0(\infty)=a_IL^{-1}+b_IL^{-2}.\label{eq:jfim}
  \end{gather}
  The dotted curves in Figs.~\ref{fig:fisher} represent the fit results and the values for the coefficients
  $a_{R/I}$ and $b_{R/I}$ are given in Table~\ref{tab:jf}. We find that the coefficient $|b_{R}|$ is roughly ten times larger than the coefficient $|a_{R}|$, which means the $L^{-1}$ and $L^{-2}$ terms give comparable contributions to
  $\text{Re}\,\kappa_0(L)$. On the other hand, the $a_{I}L^{-1}$ contribution is dominant in $\text{Im}\,\kappa_0(L)$.
These observations assure that the scaling analysis of the imaginary part is more reliable 
than the real one avoiding the possible subleading contaminations. 
In conclusion, the Fisher zero analysis indicates that the phase transition belongs to the Ising
  universality class.

  \begin{table}
    \centering
    \caption{Results for the finite size scaling analysis on both the real and imaginary parts of the Fisher
    zero.}
    \label{tab:fisher}
    \begin{ruledtabular}
    \begin{tabular}{ccccc}\toprule
      &$\nu$&$\text{Re/Im}\,\kappa_0(\infty)$&fit range&$\chi^2/{\rm d.o.f}$\\ \hline
      $\text{Re}\,\kappa_0$&$0.779(23)$&$0.241593(41)$&$L\in [16,128]$&$0.38$ \\
      $\text{Im}\,\kappa_0$&$1.030(14)$&$-0.000002(68)$&$L\in [16,128]$&$0.53$ \\ \bottomrule
    \end{tabular}
    \end{ruledtabular}
  \end{table}

  \begin{table}
    \centering
    \caption{Fit results including the subleading finite size contribution. The fit ranges are the same as
      in Table~\ref{tab:fisher}.}
    \label{tab:jf}
    \begin{ruledtabular}
    \begin{tabular}{ccccc}\toprule
      &$\text{Re/Im}\,\kappa_0(\infty)$&$a_{R/I}$&$b_{R/I}$&$\chi^2/{\rm d.o.f}$\\ \hline
      $\text{Re}\,\kappa_0$&$0.241466(37)$&$0.0636(39)$&$0.520(67)$&$0.23$ \\
      $\text{Im}\,\kappa_0$&$0.000075(37)$&$0.2578(39)$&$-0.138(67)$&$0.48$ \\ \bottomrule
    \end{tabular}
    \end{ruledtabular}
  \end{table}

  \subsection{Lee-Yang zero analysis}
  $\theta$ is regarded as an external field parameter. Scaling behavior of partition
  function zeros in the complex $\theta$ plane should be different from Eqs.~\eqref{eq:rek} and \eqref{eq:imk}.
  It is controlled by another critical exponent at the critical end point $\kappa_c$:
  \begin{equation}
    \text{Im}\,\theta_0(L)-\text{Im}\,\theta_0(\infty)\propto L^{-\left(\frac{2\delta}{1+\delta}\right)}
    =L^{-\left(\frac{2\nu-\beta}{\nu}\right)}\label{eq:ly2nd},
  \end{equation}
  where $\theta_0(L)$ is the position of a partition function zero in the complex $\theta$ plane for the
  lattice size $L$ and $\text{Im}\,\theta_0(\infty)$ is expected to be zero. 
$\delta$ and $\beta$ are the critical isotherm exponent and the critical exponent for
  magnetization, respectively. In case of the first order phase transition at $\kappa < \kappa_c$, $\theta_0(L)$ should scale in inverse proportion to the 2D lattice volume:
  \begin{equation}
    \text{Im}\,\theta_0(L)-\text{Im}\,\theta_0(\infty)\propto L^{-2}\label{eq:ly1st},
  \end{equation}
  where $\text{Im}\,\theta_0(\infty)$ should be zero. We may find $\text{Im}\,\theta_0(\infty)\ne 0$ at $\kappa > \kappa_c$, where no phase transition is expected. Note that $\text{Re}\,\theta_0(L)$ is always fixed at $\pi$ so that 
all the Lee-Yang
  zeros reside on the line $\text{Re}\,\theta=\pi$. 

  \begin{figure}
    \centering
    \includegraphics[width=80mm]{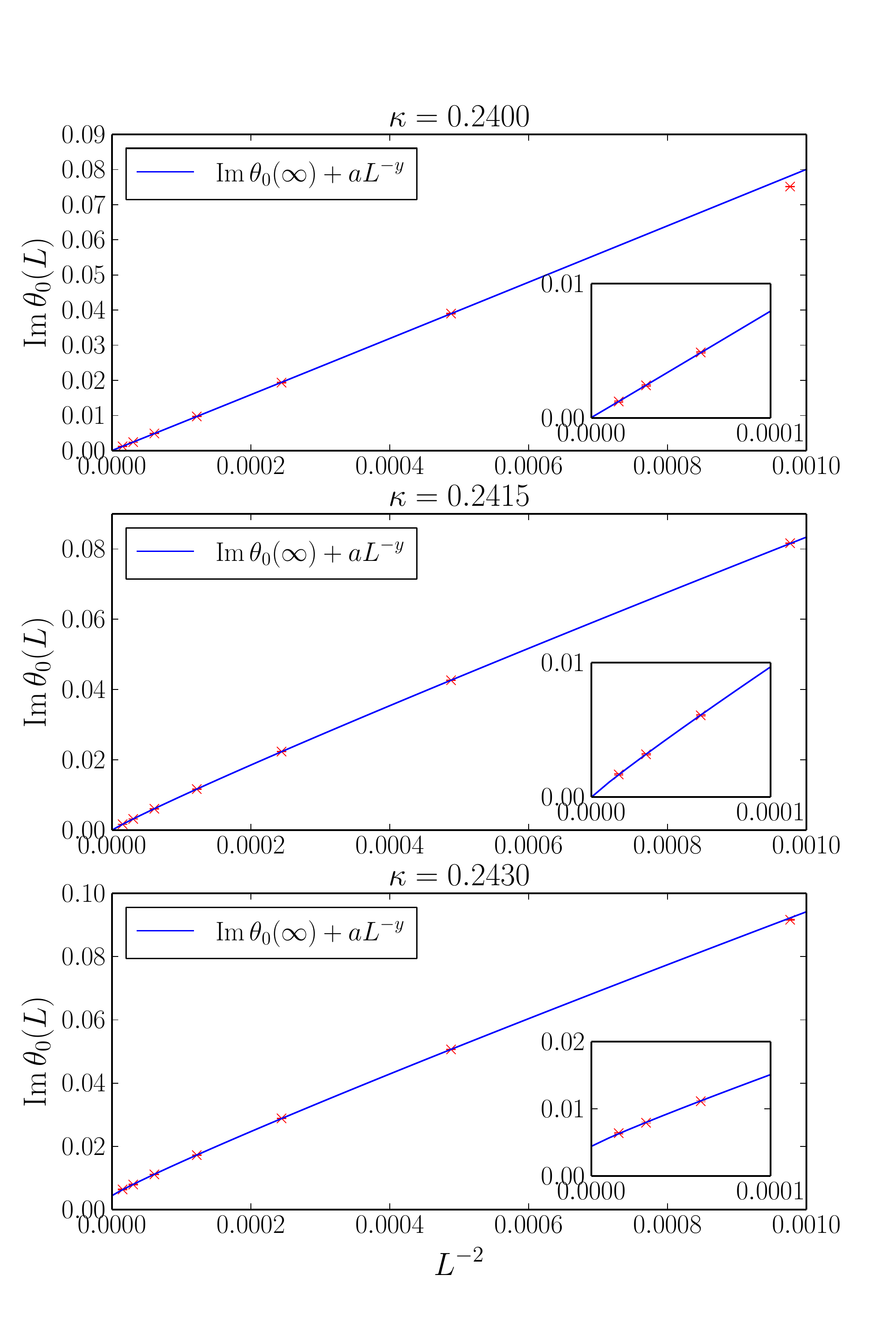}
    \caption{Imaginary part of the Lee-Yang zero for $\kappa=0.2400$ (top), $\kappa=0.2415$ (middle) and
      $\kappa=0.2430$ (bottom) as a function of $L^{-2}$ at $\beta=10.0$. Solid curves represent the fit
      results with $\text{Im}\,\theta_0(L)=\text{Im}\,\theta_0(\infty)+aL^{-y}$. Intercepts are magnified
      in small windows.}
    \label{fig:ly}
  \end{figure}

  In Fig.~\ref{fig:ly}, we present the scaling behavior of $\text{Im}\,\theta_0(L)$ at $\kappa=0.2400$, 0.2415 and 0.2430 as a function of
  $L^{-2}$. We expect $\kappa=0.2415$ is (almost) on the critical end point based on the Fisher zero analysis in the previous section. 
The solid curves denote the fit results with $\text{Im}\,\theta_0(L)=\text{Im}\, \theta_0(\infty)
  +aL^{-y}$. We choose $L\in [32\sqrt{2},256]$ for the fit range. Numerical values for the fit results 
of $\text{Im}\, \theta_0(\infty)$ and $y$ are presented in Table~\ref{tab:ly}. 
For $\kappa=0.2400$, which is smaller than
  $\kappa_c$, the inverse dependence on $L^{2}$ with $\text{Im}\, \theta_0(\infty)=0$ 
is clearly observed. It leads us to the conclusion that
  there is a first order phase transition. On the other hand, $\text{Im}\,\theta_0(\infty)$ shows 
clear deviation from zero
  at $\kappa=0.2430 > \kappa_c$, which means there is no phase transition as expected.
  For $\kappa=0.2415\approx \kappa_c$, the fit results give $y=1.869(10)$ and 
$\text{Im}\, \theta_0(\infty)=-0.000016(64)$. If there occurs a second order phase transition belonging to the Ising universality class, the critical exponent should be $y=1.875$ with
  $\delta=15$, $\beta=0.125$ and $\nu=1$ in Eq.~\eqref{eq:ly2nd}, which is consistent with our result within the error bar.
  The Lee-Yang zero analysis indicate that the phase transition at $\kappa_c$ belongs to the
  Ising universality class. It also
  agrees with the conclusion of the Fisher zero analysis.

  \begin{table}
    \centering
    \caption{Results for the finite size scaling analysis on the imaginary part of the Lee-Yang zero.}
    \label{tab:ly}
    \begin{ruledtabular}
    \begin{tabular}{ccccc}\toprule
      $\kappa$&$y$&$\text{Im}\,\theta_0(\infty)$&fit range&$\chi^2/{\rm d.o.f}$\\ \hline
      $0.2400$&$2.009(12)$&$0.000034(59)$&$L\in [32\sqrt{2},256]$&$0.65$ \\
      $0.2415$&$1.869(10)$&$-0.000016(64)$&$L\in [32\sqrt{2},256]$&$0.41$ \\
      $0.2430$&$1.850(15)$&$0.00442(12)$&$L\in [32\sqrt{2},256]$&$0.78$ \\ \bottomrule
    \end{tabular}
    \end{ruledtabular}
  \end{table}

  \section{Summary and Outlook}
  \label{sec:sum}
  We have investigated the phase structure of the lattice Schwinger model with the $\theta$ term through the
  Lee-Yang and Fisher zero analyses using the GTRG method.
  We have succeeded in reproducing the expected phase structure at $\theta=\pi$. When $\kappa$ is small,
  namely, the fermion mass is large, there exists a first order phase transition and it terminates at
  $\kappa_c$ which has a second order phase transition belonging to the Ising universality class.
 It is shown that the GTRG is applicable to the physical system with the $\theta$ term whose action is a complex number.

Extrapolation of the critical endpoint to the continuum limit was already studied by the Hamiltonian formulation with the staggered fermion employing the density matrix renormalization
group approach~\cite{Byrnes:2002nv}. It is interesting to check whether different formulations yield a consistent result. However, the naive Wilson fermion employed in this work is not suited for the detailed study of the continuum extrapolation. We will revisit it after the extension of our formulation to the $O(a)$-improved fermion action.

  \begin{acknowledgments}
    Numerical calculations for the present work were mainly carried out using the computer facilities at Research Institute for
    Information Technology, Kyushu University.
    Part of the computation was performed on the RIKEN Integrated Cluster of Clusters and the COMA (PACS-IX) computer under the ``Collaborative Interdisciplinary Program'' of Center for Computational Sciences, University of Tsukuba.
  \end{acknowledgments}

\end{document}